# ELECTRONIC STRUCTURES OF FREE-STANDING NANOWIRES MADE FROM INDIRECT BANDGAP SEMICONDUCTOR GALLIUM PHOSPHIDE

Gaohua Liao[1,2], Ning Luo[2], Ke-Qiu Chen[1,*], H. Q. Xu[2,3,*]

[1] Department of Applied Physics, School of Physics and Electronics, Hunan University, Changsha 410082, China

[2] Key Laboratory for the Physics and Chemistry of Nanodevices and Department of Electronics, Peking University, Beijing 100871, China

[3] Division of Solid State Physics, Lund University, Box 118, S-221 00 Lund, Sweden

* Emails: hqxu@pku.edu.cn (H. Q. Xu); keqiuchen@hnu.edu.cn (Ke-Qiu Chen)

**We present a theoretical study of the electronic structures of freestanding nanowires made from gallium phosphide (GaP)—a III-V semiconductor with an indirect bulk bandgap. We consider [001]-oriented GaP nanowires with square and rectangular cross sections, and [111]-oriented GaP nanowires with hexagonal cross sections. Based on tight binding models, both the band structures and wave functions of the nanowires are calculated. For the [001]-oriented GaP nanowires, the bands show anti-crossing structures, while the bands of the [111]-oriented nanowires display crossing structures. Two minima are observed in the conduction bands, while the maximum of the valence bands is always at the Γ-point. Using double group theory, we analyze the symmetry properties of the lowest conduction band states and highest valence band states of GaP nanowires with different sizes and directions. The band state wave functions of the lowest conduction bands and the highest valence bands of the nanowires are evaluated by spatial probability distributions. For practical use, we fit the confinement energies of the electrons and holes in the nanowires to obtain an empirical formula.**

## Introduction

Gallium phosphide (GaP) is an important Group III-V semiconductor material with a wide band gap of 2.272 eV at 300 K, making it attractive for use in optical devices, light-emitting diodes, and photoelectrochemical cells.[1-3] In recent years, GaP nanostructures have received considerable attention because of their potential applications in miniature optical and optoelectronic devices with enhanced performances and functyionalities.[4-7] GaP nanowires are often grown by the vapour–liquid–solid growth mechanism, which can provide nanowires with different crystal structures.[8] Typical dimensions of nanowires are diameters of the order of a few to one hundred nanometers and lengths of several micrometers. The techniques used to grow nanowires have been well developed during the last two decades. It is possible to tailor both the crystal phase and geometry of nanowire arrays, allowing control of the optoelectronic characteristics of semiconductors.[4-6] In addition, nanowires, nanowire heterostructures and superlattices can be doped in a controlled manner.[9-11] For further development and optimization of devices containing nanowires, it is important to clearly understand the electronic structures of the nanowires along common crystallographic directions such as the [001] and [111] directions. Previously, theoretical methods, such as first-principles methods,[12-15] the **k·p** method,[16-18] pseudopotential methods[19] and tight-binding methods[20-29], have been used to study semiconductor nanowires. However, first-principles calculations are impractical for unit cells containing tens of thousands of atoms. Although the **k·p** method can generally give a solution for the band states around the Γ-point, it will sometimes overestimate the symmetry of a system and could not be applied for a semiconductor system with an indirect bandgap. Compared with other approaches, tight-binding methods satisfy the accuracy condition and have the advantage of atomistic description at the nanometer scale. Tight-binding methods are capable to calculate the electronic structures of nanowires made from a semiconductor with an indirect bandgap and can easily be used to accurately treat the band structures of the nanowires with diameters in the range of a few nanometers to more than one hundred nanometers in the whole Brillouin zone.[30-33]

In this paper, a tight-binding formalism that considers $sp^3s^*$ nearest-neighbour and spin–orbit interactions is used to calculate the electronic structures of [001]- and [111]-oriented GaP nanowires including their energy bands and wave functions. Bulk GaP is an indirect bandgap material and the minima of its conduction bands are located near the X-point and the lowest set of conduction bands shows a camel back structure.[24,34-36] The bands of nanostructures are more complex in terms of band folding and quantum confinement compared with those of the bulk material. It is important to choose a suitable set of available parameters to obtain reliable results, otherwise along some crystallographic directions, the band structures of GaP nanowires may show direction-dependent bandgap characteristics even for nanowires of larger sizes.[24,32,37] We take parameters from Ref. 34 which were obtained by fitting to the anisotropic hole masses at the Γ-point, and the electron masses at the Γ-point and the X-point with the properties at the Γ-point and band edges at the X- and L-point being giving the highest weights in the fitting procedure.[34,38] We examine band edge energies, and a function of the lateral size of these energies is fitted to a simple a formula that allows the quantization energies in the nanowires to be quickly estimated. In addition, we calculate the wave functions of the band states at the minimum of the conduction band and maximum of the valence band and discuss their spatial distribution propoerties.

### Results

#### Band structures of [001]-oriented GaP nanowires

In this section, the band structures of [001]-oriented GaP nanowires with square and rectangular cross sections of different sizes are calculated and their symmetry properties are investigated. For a [001]-oriented GaP nanowire with a square (rectangular) cross section, the crystallographic structure is symmetric under the operations of the $D_{2d}$ ($C_{2v}$) point group.[39,40] At the Γ-point (wave vector k=0), the band-structure Hamiltonian of the nanowire has the same symmetry as its crystallographic structure. When considering the symmetry of the system at a wavevector away from the Γ-point ($k \neq 0$), only the symmetry operations that leave the wavevector k unchanged are allowed.[41] In the case of a nanowire with a square cross section, the band-structure Hamiltonian of the nanowire is symmetric under the operations of the $C_{2v}$ point group when $k \neq 0$. Because of the inclusion of the spin–orbit interaction in the Hamiltonian, the symmetry properties of the electronic structure need to be characterised by the corresponding double point group. The $D_{2d}$ double point group has two doubly degenerate double-valued irreducible representations, $Γ_6$ and $Γ_7$, while the $C_{2v}$ double point group has only one double-valued irreducible representation, $Γ_5$.

Figure 1(a)-(d) show the band structures of [001]-oriented GaP nanowires with a square cross section with lateral sizes of 11.6×11.6 and 23.1×23.1 nm$^2$, while Fig. 1(e)-(h) present the band structures of [001]-oriented GaP nanowires with a rectangular cross section with lateral sizes of 23.1×11.6 and 34.7×11.6 nm$^2$. The valence bands of GaP nanowires with square cross sections are labelled by the symmetries of their states at the Γ-point in correspondence with the irreducible representations $Γ_6$ and $Γ_7$ of the $D_{2d}$ double point group. All the bands of the GaP nanowires with rectangular cross sections are $Γ_5$-symmetric and doubly degenerate. The band gaps of the GaP nanowires increased as their lateral size decreased because of quantum confinement. However, unlike InAs, InP, InSb and GaSb nanowires whose bulk band gaps are direct and show simple parabolic dispersions around the Γ-point, the lowest conduction bands of the GaP nanowires, whose bulk band gap is indirect, show more complex structures. Figure 1(a) and (c) show that the conduction bands of the nanowires are dense and two minima appear; one around the Γ-point (k=0) and the other around $k = 0.43\pi/a$. Magnified views of the lowest conduction bands around the two minima are provided in Fig. 2. Figure 2(a) and (c) indicate that the minimum of the conduction band that is slightly away from the Γ-point is about 1.5 meV bigger than the minimum around $k = 0.43\pi/a$. At the Γ-point, the four lowest conduction bands are hardly distinguishable and nearly eight degenerate. When the wave vector moves away from the Γ-point, these bands develop into two nearly degenerate separated groups of four. The next four lowest conduction bands show the same behaviour around the Γ-point. With increasing lateral size of the nanowires, the bands become closer. However, Fig. 2(b) reveals that the four lowest conduction bands around $k = 0.43\pi/a$ show simple parabolic dispersion relations and the band states all possessed $Γ_5$ symmetries.

The valence bands of the [001]-oriented GaP nanowires also show complex structures. In contrast to other [001]-oriented nanowires such as InAs and InP direct-gap semiconductor nanowires with valence bands with two maxima away from the Γ-point, one maximum of the valence bands of the GaP nanowires is always at the Γ-point, and the other maximum of the top valence band is much smaller than that at the Γ-point. When the cross section of a [001]-oriented nanowire is rectangular, the band-structure Hamiltonians of these nanowires are symmetric under the operations of the $C_{2v}$ point group. Thus, all the bands are doubly degenerate and $\Gamma_5$ symmetric, and no band crossings occur in the band structures. Compared with the band structures of nanowires with square cross sections, their conduction bands show similar structures except that the bands are well separated energetically. The valence bands of the nanowires with rectangular cross sections tend to show much simpler dispersions. In particular, the top valence bands of a GaP nanowire with a rectangular cross section become more parabolic around the Γ-point. Figure 1(f) and (h) show that the top valence bands of a GaP nanowire with a rectangular cross section become more parabolic around the Γ-point with increasing aspect ratio $d_1/d_2$, and the other maximum away from the Γ-point disappears.

To provide useful guidance for experimentalists to estimate band edge states, we fitted the edge energies of the lowest conduction band and topmost valence band, $E_c(d)$ and $E_v(d)$, respectively, of the [001]-oriented GaP nanowires using the following expression:

$$\Delta E_\alpha(d) = E_\alpha(d) - E_\alpha(\infty) = \frac{1}{q_1 d^2 + q_2 d + q_3} + q_4 , \qquad (1)$$

where $q_1$, $q_2$, $q_3$ and $q_4$ are fitting parameters, and $E_\alpha(\infty) = E_c$ or $E_v$, which are the band-edge energies of bulk GaP taken from Ref.34. Figure 3 shows the quantum confinement effect on the conduction and valence band states of GaP nanowires with square and rectangular cross sections. The quantum confinement energy of the electrons at the conduction band edge is very strong. For example, the quantum confinement energy of the electrons in the conduction bands of the [001]-oriented GaP nanowire with a square cross section with $d$~20 nm is only about 0.95 meV. When $d$ is increased to 35 nm, the electron quantisation energy at the conduction band edge is only 0.44 meV. The quantum confinement energies of the holes in the valence bands of the [001]-oriented GaP nanowires are generally larger than the corresponding quantum confinement energies of electrons at the conduction band edge of the nanowires. For example, for nanowires with $d$~20 nm, the quantum confinement energy of holes is about −5.11 meV, and when $d$~35 nm, the quantum confinement energy of holes is about −1.73 meV, which is still larger than the quantum confinement energy of electrons at 20 nm. Similar quantum confinement properties are also found for the [001]-oriented GaP nanowires with rectangular cross sections. Nevertheless, at the limit of large values of $d_2$, the quantum confinement energies of electrons and holes at the band edges of nanowires with a rectangular cross section approach their respective constant values of $q_4$ listed in Table 1.

**Wave functions of [001]-oriented GaP nanowires**

Representative wave functions of the band states of [001]-oriented GaP nanowires are presented in Fig. 4 and 5. All the band states are doubly degenerate and the wave function of each spin-degenerate state has an identical spatial distribution at the Γ-point (k=0). However, when *k* moves away from the Γ-point, the two spin-degenerate states may have different spatial distributions. To calculate the wave function distribution, we separated the doubly degenerate band states with different symmetries using projection operation $P_{kk}^{(j)}$. This allowed us to obtain the probability distribution on a (001) layer of Ga cations, whose value at each cation site is the sum of the squared amplitudes of all the atomic orbital components of the cation site and all the normalised probability distributions, scaled against its maximum value. As mentioned earlier, the lowest conduction band appears around $k = 0.43\pi/a$ and the highest valence band is always at the Γ-point (see Fig. 1). Figure 4 depicts the wave functions of the four lowest conduction band states (at $k = 0.43\pi/a$) and the four highest valence band states (at k=0) of [001]-oriented GaP nanowires with a square cross section with *d*=11.6 nm. The lowest conduction band states of the [001]-oriented GaP nanowire possess an *s*-like distribution. The wave functions of the second- and third-lowest conduction band states show probability distributions that have a doughnut shape, which is consistent with the fact that these two band states are nearly degenerate in energy. The fourth conduction band states show four *s*-like probability distributions at the corners of the square cross sections. As for the four highest valence band states, the first- and third-highest valence band states (at the Γ-point) have similar probability distributions to that of the lowest conduction band of the nanowire ($k \approx 0.43\pi/a$). The second-highest valence band states show a probability distribution resembling a doughnut. The fourth valence band states show four-fold symmetric probability distribution patterns and are more localised to the inner region of the nanowire.

Figure 5 presents the wave functions of the four lowest conduction bands (at $k \approx 0.43\pi/a$) and four highest valence band states (at the Γ-point) of a [001]-oriented GaP nanowire with a rectangular cross section of 23.1×11.6 nm². The lowest conduction band states in this GaP nanowire show regular probability distribution patterns with one, two and three maxima present, except for the fourth-lowest conduction band state, which displays probability distribution patterns with two maxima along the short side. With increasing $d_1/d_2$, the fourth-lowest conduction band states will present a distribution with four maxima. Figure 5 reveals that for the four highest valence band states in the GaP nanowire with a rectangular cross section, the highest valence band state is also *s*-like. The wave function of the second-highest valence band state retains some doughnut shape and is more elongated. The wave function of the third valence band state has a strong contribution in the center, and the fourth resembles the fourth valence band of the square nanowire.

**Band structures of [111]-oriented GaP nanowires**

In this section, the band structures of [111]-oriented GaP nanowires with hexagonal cross sections of different sizes are calculated and their symmetry properties are investigated. In a [111]-oriented nanowire with a hexagonal cross section, the crystallographic structure is symmetric under the operations of the $C_{3v}$ point group. Again, because of the inclusion of spin–orbit interactions in the Hamiltonian, the symmetry properties of the electronic structure need to be characterised by the corresponding double point group. The $C_{3v}$ double point group has two single-valued irreducible representations, $\Gamma_5$ and $\Gamma_6$, and one doubly degenerate double-valued irreducible representation, $\Gamma_4$.

Figure 6 shows the band structures of [111]-oriented GaP nanowires with hexagonal cross sections with lateral sizes of 11.1 and 22.3 nm. Figure 6(a) reveals that the conduction bands are dense and there are two closed minima in the lowest conduction band at $k=0.412\pi/(\sqrt{3}a)$ and $k=0.456\pi/(\sqrt{3}a)$. The crossing of two groups of bands occurs near the Γ-point; the energies of one group tend to decrease as the wave vector $k$ moves away from the Γ-point, and the other group behaves in the opposite manner. At the Γ-point, all the band states are doubly degenerate. For the [111]-oriented nanowire with a lateral size of $d=$ 11.1 nm, the symmetries of the three lowest conduction bands are $\Gamma_5$ ($\Gamma_6$), $\Gamma_4$ and $\Gamma_4$ with ascending energy. When $k$ moves away from the Γ-point, $\Gamma_5$ ($\Gamma_6$) states would split into two single degenerate states, the upper band with $\Gamma_5$ symmetry and the lower with $\Gamma_6$ symmetry. The two closed $\Gamma_4$ bands also separate, thus developing into two groups of three nearly degenerate bands. As the lateral size $d$ increases, the lowest three bands move down because of the weakened quantum confinement. Interestingly, the two lowest conduction bands with $\Gamma_4$ symmetry move to the bottom of the conduction band. Magnified views of the lowest conduction bands around the two minima are presented in Fig. 7 along with their band state symmetries. Figure 7(a) ($d=$11.1 nm) reveals that the lowest conduction bands appear as two closed minima not at the Γ-point, and the two minima possess different symmetry properties. The conduction bands around the two minima are nearly triply degenerate and show crossing characteristics. When $d$ increases to 22.3 nm, the conduction bands get denser, and the energies of the two lowest minima decrease and get closer. As we mentioned above, $\Gamma_4$ representation is a double-degenerate double-valued irreducible representations, when they get closer, anti-crossing characteristics appear to avoid exceeding the dimension of $\Gamma_4$ irreducible representation.

For practical use, we also fitted the edge energies of the lowest conduction band and topmost valence band of the [111]-oriented GaP nanowires with hexagonal cross sections using formula (1). The fitting parameters are listed in Table 1 and the data and the fitting results are presented in Fig. 8. The quantum confinement energy of electrons at the conduction band edge of the [111]-oriented GaP nanowire with a hexagonal cross section of $d$~20 nm is 1.82 meV. When the $d$ increases to ~35 nm, the quantum confinement energy of electrons decreases to 0.73 meV. The quantum confinement energies of holes in the [111]-oriented GaP nanowires with $d$~20 and 35 nm are about −5.78 and −2.03 meV, respectively.

**Wave functions of [111]-oriented GaP nanowires**

We also calculated the wave functions of [111]-oriented GaP nanowires. The representative results are shown in Fig. 9 and 10. Here, a wave function is represented by the probability distribution on the (111) plane of Ga cations with the probability at each cation site calculated as the sum of the squared amplitudes of all the atomic orbital components on that site scaled against the maximum value in each graph. These figures present the lowest conduction band states and few highest valence band states. As we mentioned above, at the Γ-point (k=0), all the band states are doubly degenerate, while when $k \neq 0$, the $\Gamma_4$ bands still are doubly degenerate but $\Gamma_5$ and $\Gamma_6$, which are degenerate at the Γ−point, split into two non-degenerate bands. The lowest conduction band of a [111]-oriented GaP nanowire has two closed minima not at the Γ-point. At the Γ-point, only one of the spin-degenerate wave functions is presented because the other one has an identical spatial component. All the spin-degenerate wave functions are presented for cases where the wave vector is not at the Γ-point. Figures 9(a) and 9(b) depict the lowest conduction band states at $k = 0.412\pi/(\sqrt{3}a)$ and $k = 0.456\pi/(\sqrt{3}a)$, respectively, for a [111]-oriented nanowire with a lateral size $d$= 11.1 nm, and Figs. 9(c) and 9(d) present the lowest conduction band states of a nanowire with $d$=22.3 nm. The state at $k = 0.412\pi/(\sqrt{3}a)$ (a $\Gamma_4$ state) shows an s-like wave function probability distribution but the superposed states lead to the formation of interference patterns. Another lowest conduction band state at $k = 0.456\pi/(\sqrt{3}a)$ (a $\Gamma_6$ state) shows a similar distribution and more obvious interference patterns. Similar patterns are observed when $d$ is increased to 22.3 nm (Figs. 9(c) and 9(d)).

The highest valence band states are located at the Γ-point. Figure 10(a)-(d) show the four highest valence band states of [111]-oriented GaP nanowires with a hexagonal cross section of $d$=11.1 nm. The first- and fourth-highest valence band states show s-like distributions, while the second- and third-highest valence band states display doughnut distributions. When the lateral size is increased to $d$=22.3 nm, the first- and second-highest valence band states show nearly distinct distributions, while the third- and fourth-highest valence band states exchange their distributions while retaining the same symmetric order.

**Discussion**

We performed a theoretical study of the electronic structures of free-standing zincblende [001]-oriented GaP nanowires with square and rectangular cross sections and [111]-oriented GaP nanowires with hexagonal cross sections by the atomistic $sp^3s^*$ nearest-neighbour tight-binding method considering spin–orbit interactions. The band structures and band-state wave functions of these nanowires were calculated and the symmetry properties of the band states were analysed based on their double point groups. All bands of [001]-oriented nanowires are doubly degenerate and possess $D_{2d}$ symmetry properties at the Γ-point, while the decreased symmetry leads to the bands displaying $C_{2v}$ symmetry properties when $k \neq 0$. For the [111]-oriented nanowires, all the bands are doubly degenerate at the Γ-point and some of these bands split into non-degenerate bands when k moves away from the Γ-point as a manifestation of spin splitting caused by spin–orbit interactions. The lower conduction bands of the [001]-oriented nanowires with square and rectangular cross sections displayed two maxima around the Γ-point and simple parabolic dispersion relations around $k = 0.43\pi/a$. However, the top valence bands of the [001]-oriented nanowires with square and rectangular cross sections show complex dispersion relations and band anti-crossings, and the minima are always at the Γ-point. In the case of [001]-oriented nanowires with a rectangular cross section, the valence bands tend to become more parabolic as the aspect ratio of the cross section is increased. The wave functions of the band states of [001]- and [111]-oriented GaP nanowires are presented as probability distributions on cross sections. The wave functions of the band states in the [001]-oriented nanowires with square and rectangular cross sections are characteristically different. The spatial probability distributions of the band states in the nanowires with a square cross section show interesting structures with characteristics that are more complex than the predictions of simple one-band effective mass theory. Meanwhile, the spatial probability distributions of the band states in the nanowires with rectangular cross sections show characteristics that could be well described by one-band effective mass theory. For the [111]-oriented GaP nanowires with hexagonal cross sections, the conduction band states show interference patterns because of the interactions of the band states. All the highest valence band states show $2\pi/3$-rotation symmetric probability distributions and two degenerate band states at the Γ-point with identical spatial probability distributions. Finally, the effects of quantum confinement on the band structures of the [001]- and [111]-oriented GaP nanowires were examined and an empirical formula to describe the quantisation energies of the lowest conduction band and highest valence band was developed. This formula can be used to simply estimate the enhancement of the band gaps of nanowires of different sizes caused by quantum confinement. We believe that the results presented in this work provide important information about the electronic structures of [001]- and [111]-oriented GaP nanowires and guidance for the use of these nanowires in novel nanoelectronic, optoelectronic and quantum devices.

**Method**

The nanowires considered in this paper are zincblende crystals oriented along the [001] and [111] crystallographic directions with $\{1\bar{1}0\}$ facets. Because only the translational symmetry along the growth direction is preserved in the nanowire geometry, the period of a unit cell is a in a [001]-oriented nanowire and $\sqrt{3}a$ in a [111]-oriented nanowire, where a= 0.54509 nm is the lattice constant of bulk GaP. For a [001]-oriented nanowire unit cell with a rectangular cross section, the lengths of the two sides are $d_1$ and $d_2$ (for square cross sections $d_1=d_2$). Here, $d_1$ and $d_2$ possess discrete values of $d_1 = n_1 a/\sqrt{2}$ and $d_2 = n_2 a/\sqrt{2}$, where $n_1$ and $n_2$ are positive integers (for square cross sections $n_1 = n_2$). For a unit cell of a [111]-oriented nanowire with a hexagonal cross section, the distance between two parallel sides in the hexagonal cross section is $d_1$ and the distance between two most remote corners is $d_2$. Here, $d_1$ and $d_2$ possess discrete values of $d_1 = na/\sqrt{2}$ and $d_2 = 2na/\sqrt{6}$, where $n$ is a positive integer. In the present work, without confusion, we take d=$d_2$ to represent the cross sectional size of the [111]-oriented nanowires. In our calculation, the $sp^3s^*$ nearest-neighbour tight-binding formalism was used to determine the electronic structures of the [001]- and [111]-oriented GaP zincblende nanowires considering their spin–orbit interaction. Tight-binding parameters were taken from Ref. 34. In the tight-binding formalism, Bloch sums of the form[31,42]

$$|\alpha, \nu, \mathbf{k}\rangle = N^{-1/2} \sum_{\mathbf{R}} e^{i\mathbf{k}\cdot\mathbf{R}_\nu} |\alpha, \mathbf{R}_\nu\rangle \ , \qquad (2)$$

were used as a basis, where $|\alpha, \mathbf{R}_\nu\rangle$ stands for an atomic orbital $\alpha$ at position $\mathbf{R}_\nu$, and N is the number of lattice sites. In the basis of the Bloch sums, the Hamiltonian can be written in a matrix form with matrix elements given by

$$\mathbf{H}_{\alpha\nu,\beta\xi}(\mathbf{k}) = \sum_{\mathbf{R}} e^{i\mathbf{k}\cdot(\mathbf{R}'_\nu - \mathbf{R}_\xi)} \langle\alpha, \mathbf{R}'_\nu | \mathbf{H} | \beta, \mathbf{R}_\xi\rangle \ , \qquad (3)$$

$$|n, \mathbf{k}\rangle = \sum_{\alpha,\nu} c_{n,\alpha\nu} |\alpha, \nu, \mathbf{k}\rangle \ . \qquad (4)$$

To eliminate the effects of dangling bonds on the band states near the fundamental band gaps, the dangling bonds at the surface of the nanowires were passivated using hydrogen atoms. The parameters involved in the passivation using hydrogen atoms were determined by the procedure presented in Refs. 43 and 44. For further details about the calculation methods, please refer to Refs. 26 and 28.

A unit cell in a nanowire with a large lateral size consists of an extremely large number of atoms and the resulting Hamiltonian matrix becomes too large to be solved by a standard diagonalisation procedure (the largest Hamiltonian matrix in this paper is of the order of 390,000 × 390,000). In this work, the Lanczos algorithm[45] was used to solve the eigenvalues and eigenvectors of Eq. (3). To separate the doubly degenerate band states, the symmetry properties of the nanowires were exploited to project out the part of given vector $|\phi_n\rangle$ belonging to the $kth$ row of the $jth$ representation,

$$|\phi_{n,k}^{(j)}\rangle = P_{kk}^{(j)} |\phi_n\rangle \ , \tag{5}$$

where

$$P_{kk}^{(j)} = \frac{l_j}{h} \sum_R \Gamma^{(j)}(R)_{kk}^* R \ , \tag{6}$$

and $\Gamma^{(j)}(R)_{kk}$ is the $kth$ diagonal element in the $jth$ irreducible representation of the symmetric operation $R$, $l_j$ is the dimensionality of the $jth$ representation, and $h$ is the number of symmetric operations in a point group.

**Acknowledgements**

This work was supported by the National Basic Research Program of China (Grants No.2012CB932703 and No.2012CB932700) and the National Natural Science Foundation of China (Grants Nos. 91221202, 91421303, 61321001, and 11274105).

**Author contributions**

K. Q. C. and H. Q. X. conceived the research. G. H. L. performed all the calculations, G. H. L., K. Q. C. and H. Q. X. wrote the manuscript and G. H. L. prepared all the figures. All authors reviewed and discussed the manuscript.

**Additional information**

**Competing financial interests:** The authors declare no competing financial interests.

**Figure Legends**

Figure 1 | Band structures of [001]-oriented GaP nanowires with a cross section of the lateral sizes of (a) and (b) 11.6×11.6 nm$^2$, (c) and (d) 23.1 × 23.1 nm$^2$, (e) and (f) 23.1 × 11.6 nm$^2$, (g) and (h) 34.7 × 11.6 nm$^2$. The valence bands of GaP nanowires with square cross sections are labeled by the symmetries of their states at the Γ-point in correspondence with the irreducible representations, Γ$_6$ and Γ$_7$, of the $D_{2d}$ double point group. All the bands of GaP nanowires with rectangular cross sections are Γ$_5$ -symmetric and doubly degenerate.

Figure 2 | Zoom-in plots of the conduction band structures of [001]-oriented GaP nanowires with a square cross section of the lateral sizes of (a) and (b) 11.6×11.6 nm$^2$, and (c) and (d) 23.1×23.1 nm$^2$.

Figure 3 | Lowest conduction band electron and highest valence band hole confinement energies in the [001]-oriented GaP nanowires as a function of the lateral size. Panels (a) and (b) show the results for the GaP nanowires with square cross sections and panels (c) and (d) show the results for the GaP nanowires with rectangular cross sections. The calculated data are presented by symbols "∗" and the solid lines are the results of fittings based on Eq. (1) with the fitting parameters listed in Table 1. The insets show the zoom-in plots of the calculated confinement energies in the nanowires at large sizes.

Figure 4 | Wave functions of the four lowest conduction band states ( $k = 0.43\pi/a$ ) and the four highest valence band states(k=0) of the [001]-oriented GaP nanowires with a square cross section of the size 11.6×11.6 nm$^2$.

Figure 5 | Wave functions of the four lowest conduction band states( $k = 0.43\pi/a$ ) and the four highest valence band states(k=0) of the [001]-oriented GaP nanowires with a square cross section of the size 23.1×11.6 nm$^2$.

Figure 6 | Band structures of [111]-oriented GaP nanowires with a hexagonal cross section of the lateral sizes of (a) and (b) 11.1 nm, (c) and (d) 22.3 nm. The valence bands of GaP nanowires with hexagonal cross sections are labeled by the symmetries of their states at the Γ-point in correspondence with the irreducible representations, Γ$_4$ , Γ$_5$ and Γ$_6$ of the $C_{3v}$ double point group. In the figures, the order of Γ$_5$ and Γ$_6$ corresponds to a band initially has a lower energy after splitting as the wave vector k moves away from the Γ-point.

Figure 7 | Zoom-in plots of the conduction band structures of [111]-oriented GaP nanowires with a hexagonal cross section of the lateral sizes of (a) 11.1 nm, (b) 22.3

nm. The symmetry properties of band states are presented by the solid lines and symbols with different colors.

Figure 8 | Lowest conduction band electron and highest valence band hole confinement energies in the [111]-oriented GaP nanowires with a hexagonal cross section as a function of the lateral size $d$. The calculated data are presented by symbols "*" and the solid lines are the results of fittings based on Eq. (1) with the fitting parameters listed in Table 1. The insets show the zoom-in plots of the calculated confinement energies in the nanowires at large sizes.

Figure 9 | Wave functions of the lowest conduction band states at $k = 0.412\pi/(\sqrt{3}\,a)$ and $k = 0.456\pi/(\sqrt{3}\,a)$ of the [111]-oriented GaP nanowires with a hexagonal cross section of the size (a) and (b) $d$=11.1 nm and (c) and (d) $d$=22.3 nm.

Figure 10 | Wave functions of the four highest valence band states at the Γ-point (k = 0) of the [111]-oriented GaP nanowires with a hexagonal cross section of the size (a)-(e) $d$=11.1 nm and (f)-(j) $d$=22.3 nm.

**Tables**

Table 1

Parameters $q_1$, $q_2$, $q_3$, and $q_4$ in Eq. (1) obtained by fitting the equation to the calculated energies of the lowest conduction band and the highest valence band of the [001]-oriented GaP nanowires with a square cross section (labeled by subscript "squ") and with a rectangular cross section (labeled by subscript "rec") and the [111]-oriented GaP nanowires with a hexagonal cross section (labeled by subscript "hex").

| Nanowire type | Band shift (eV) | $q_1$ (eV$^{-1}$nm$^{-2}$) | $q_2$ (eV$^{-1}$nm$^{-1}$) | $q_3$ (eV$^{-1}$) | $q_4$ (eV) |
|---|---|---|---|---|---|
| [001]$_{squ}$ | $\Delta E_c$ | 2.92638 | 1.03589 | 1.50493 | |
| [001]$_{squ}$ | $\Delta E_v$ | -0.45261 | -0.89274 | -0.09389 | |
| [001]$_{rec}$ | $\Delta E_c$ | 6.18442 | 0.44825 | 3.06074 | 0.00133 |
| [001]$_{rec}$ | $\Delta E_v$ | -0.50271 | 2.13146 | -47.43079 | -0.00323 |
| [111]$_{hex}$ | $\Delta E_c$ | 1.36955 | 1.4002 | 1.19828 | |
| [111]$_{hex}$ | $\Delta E_v$ | -0.33118 | -2.07445 | 0.49631 | |

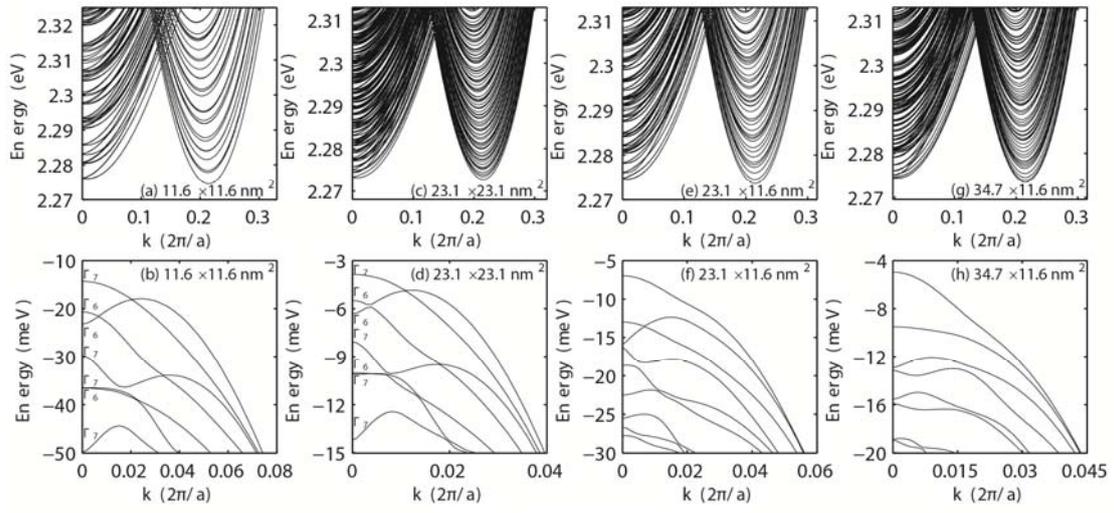

Figure 1, Gaohua Liao et al.

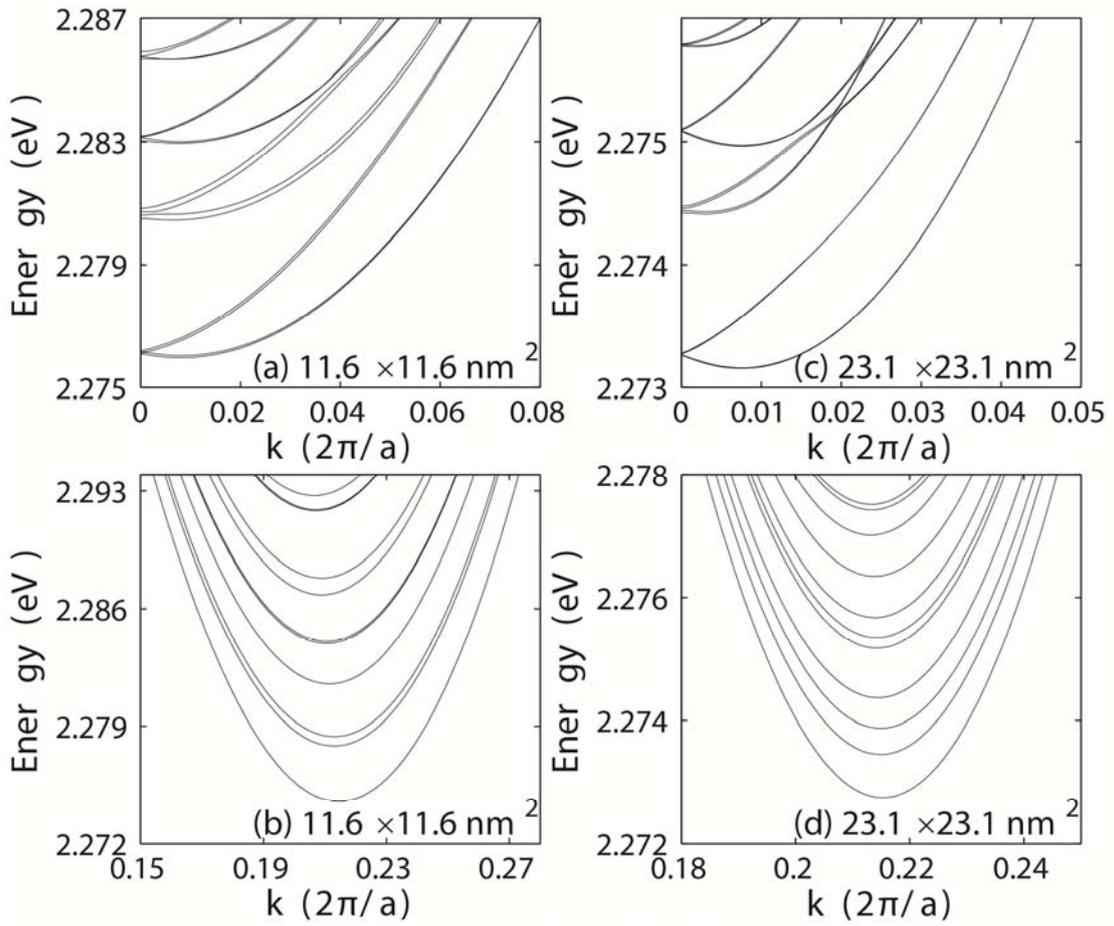

Figure 2, Gaohua Liao et al.

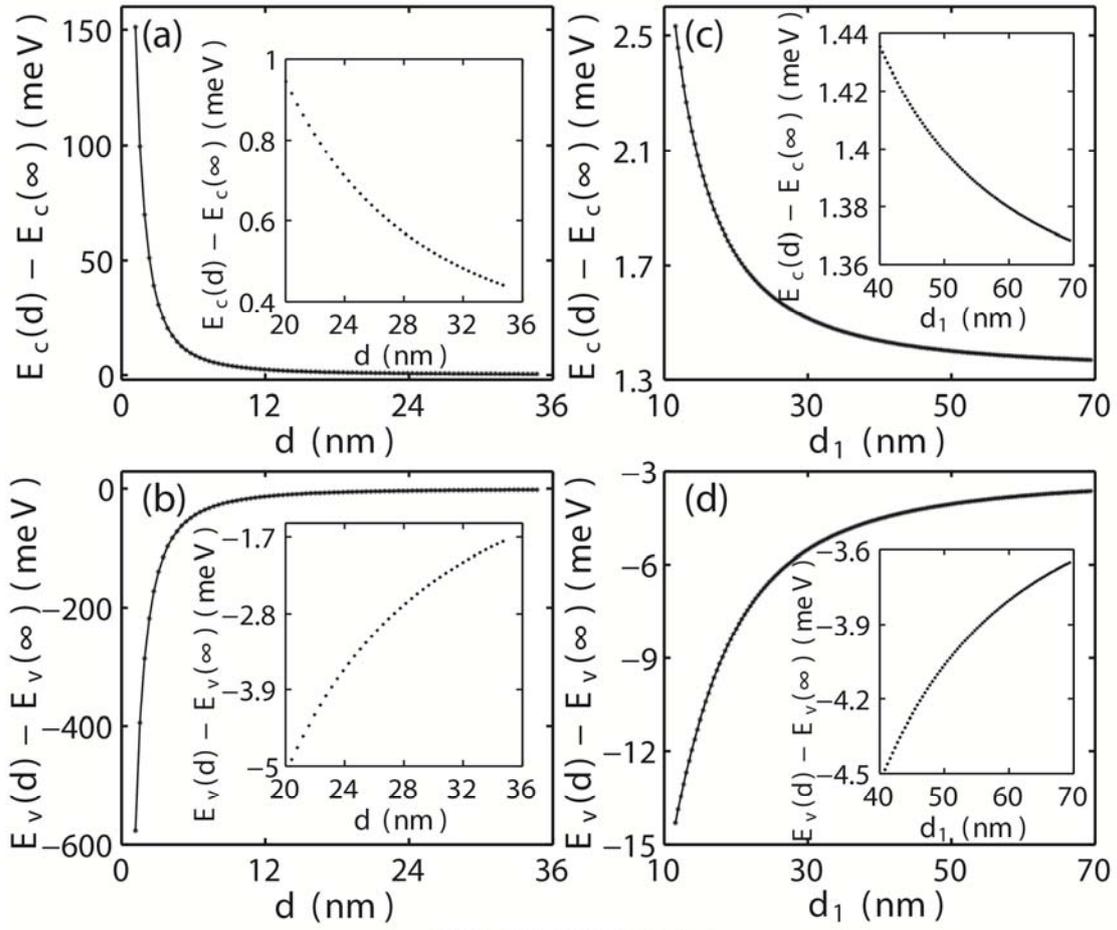

Figure 3, Gaohua Liao et al.

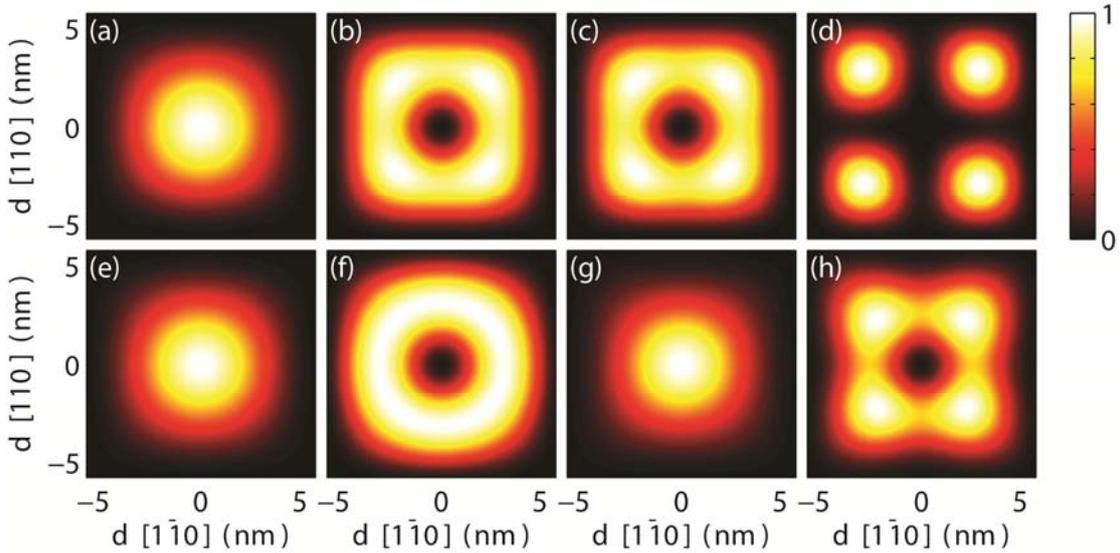

Figure 4, Gaohua Liao et al.

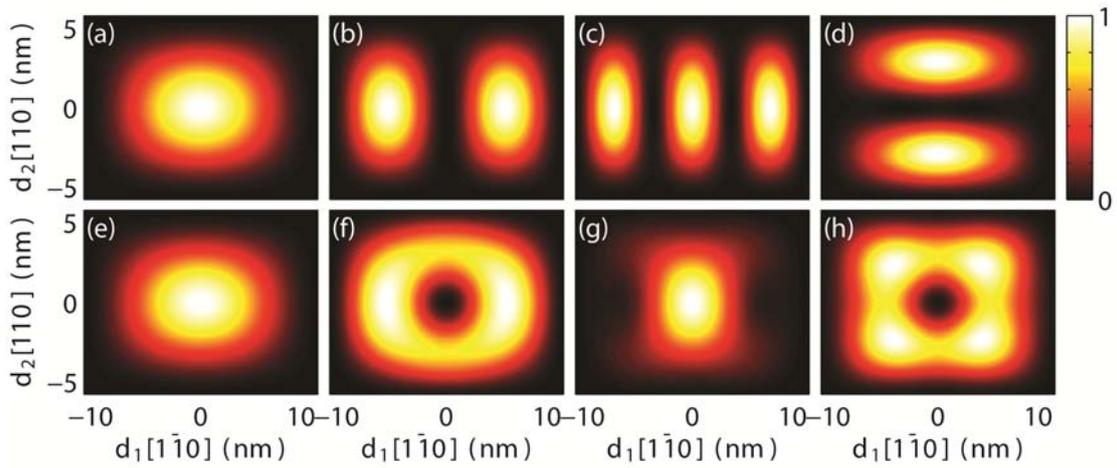

Figure 5, Gaohua Liao et al.

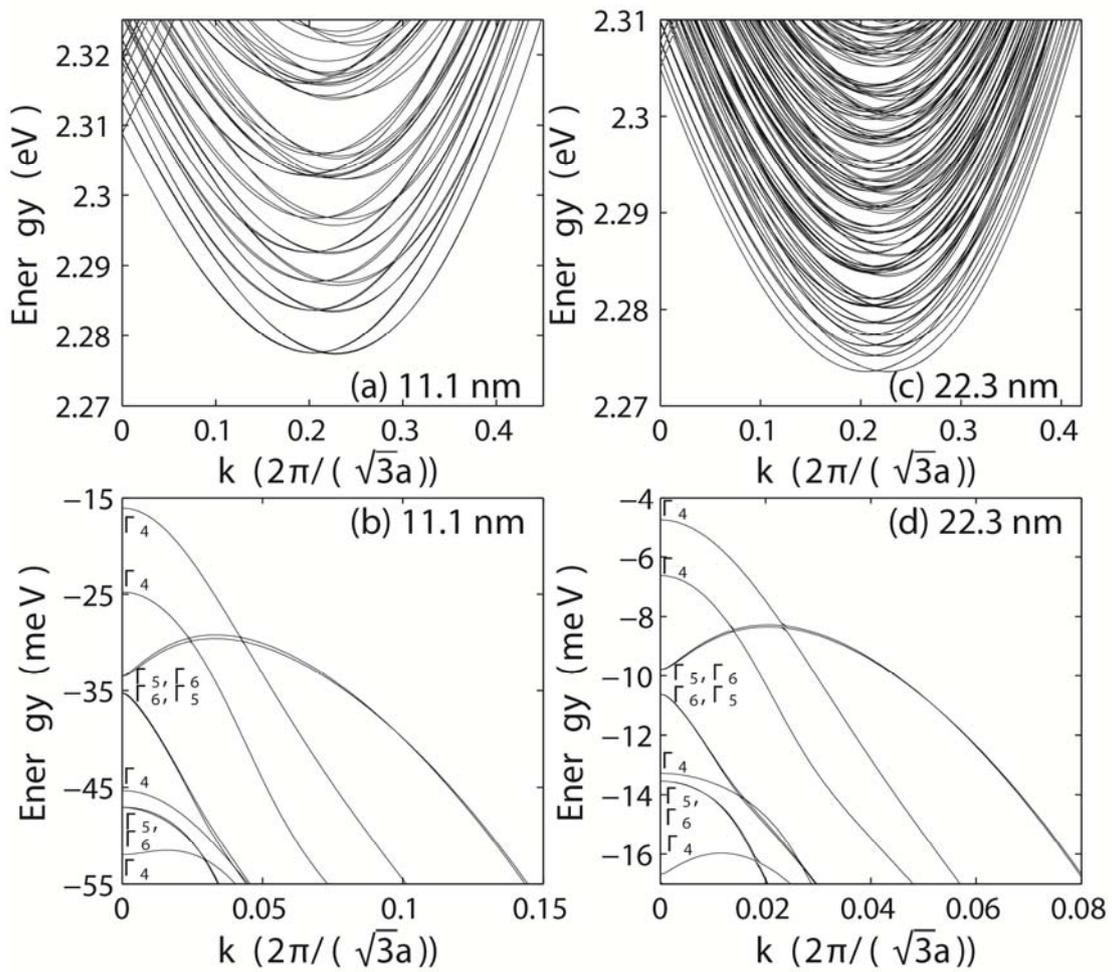

Figure 6, Gaohua Liao et al.

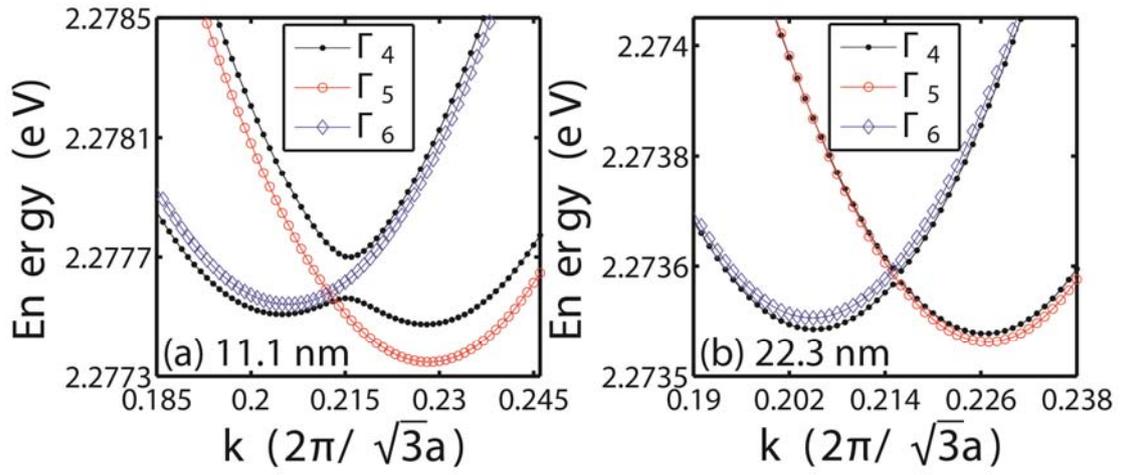

Figure 7, Gaohua Liao et al.

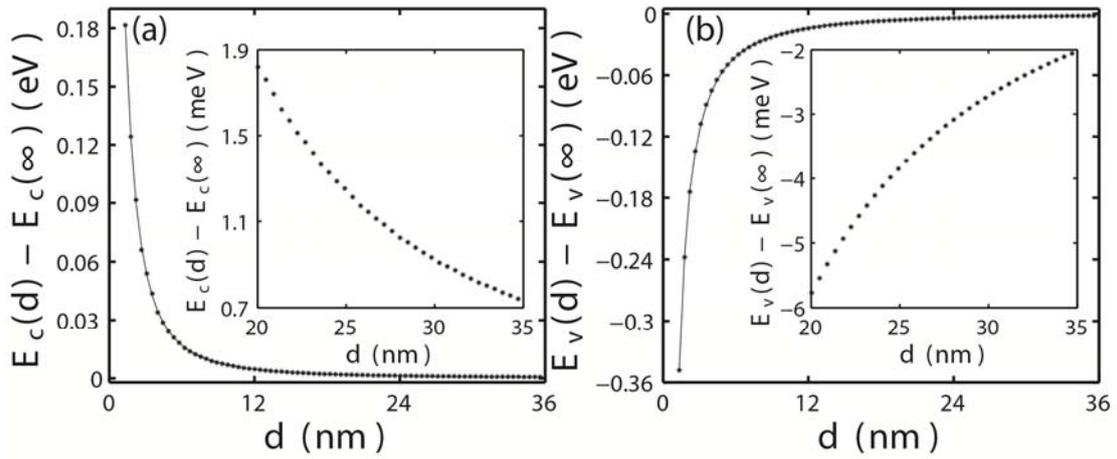

Figure 8, Gaohua Liao et al.

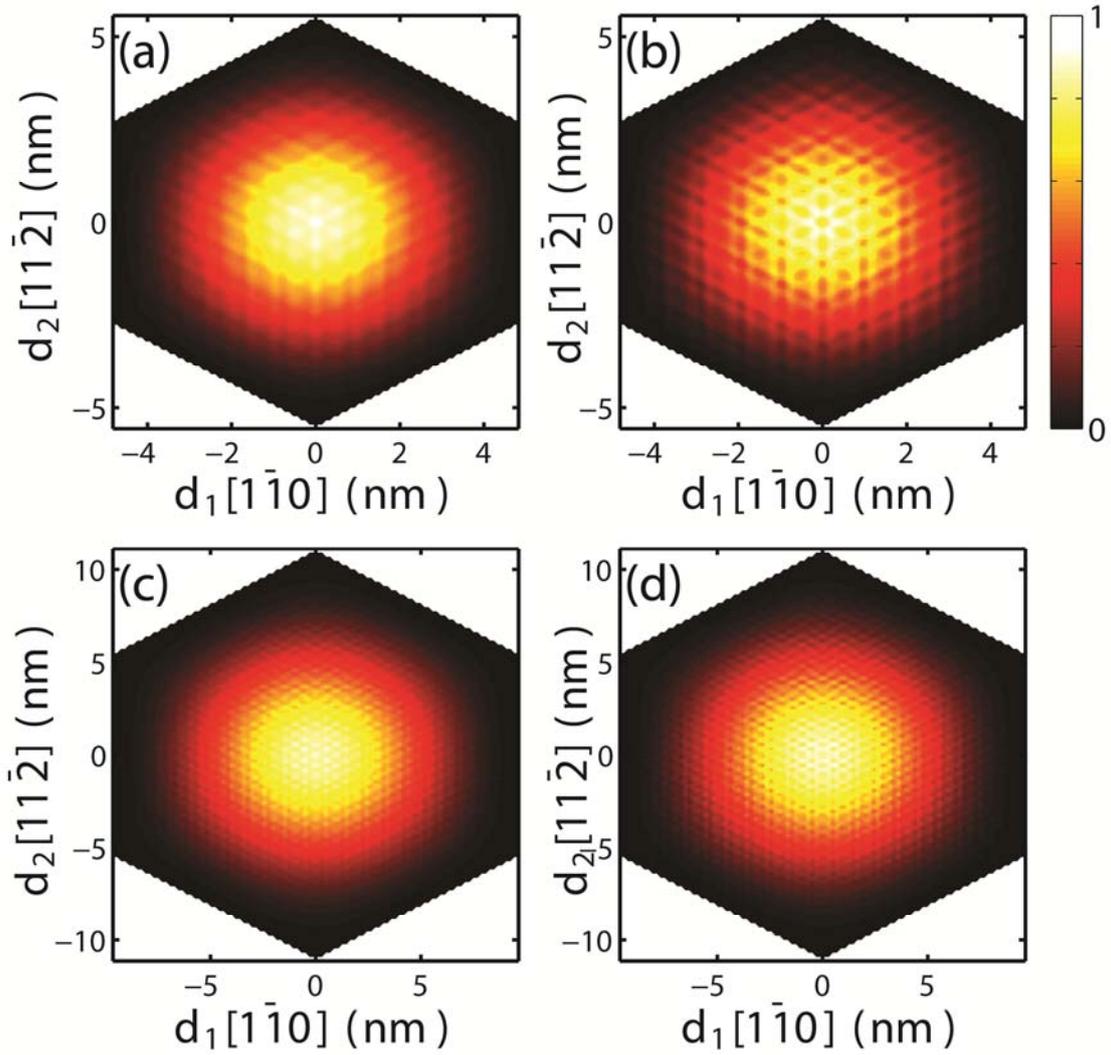

Figure 9, Gaohua Liao et al.

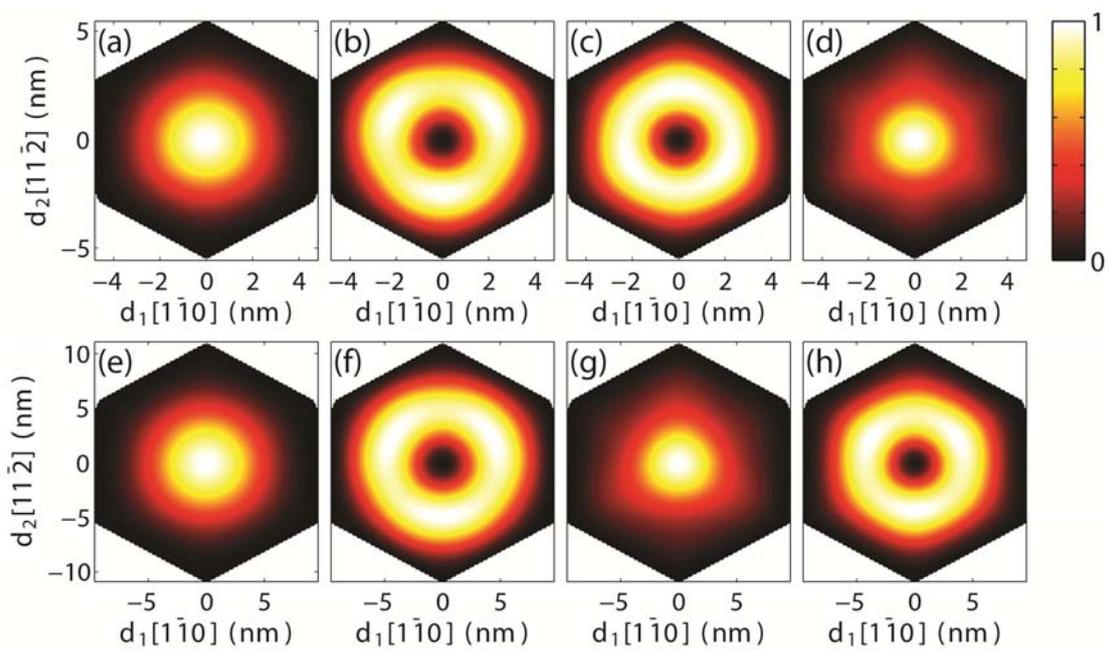

Figure 10, Gaohua Liao et al.

19